\documentclass[10pt, conference, letterpaper]{IEEEtran}
\usepackage{graphicx}
\usepackage{subfigure}
\usepackage{makecell}
\usepackage{amsmath}
\usepackage{amssymb}
\usepackage{amsthm}
\usepackage{url}
\usepackage{multirow}
\usepackage[ruled,vlined]{algorithm2e}
\usepackage{enumitem}
\usepackage{tabularx}
\usepackage[table,xcdraw]{xcolor}
\usepackage{listings}

\setitemize{leftmargin=4mm}

\usepackage{tikz}
\newcommand*\circled[1]{\tikz[baseline=(char.base)]{
            \node[shape=circle,draw,inner sep=1pt] (char) {#1};}}

\begin{document}

\author{Zili Meng$^{1,2}$, Jun Bi$^{1,2}$, Haiping Wang$^{1,2}$, Chen Sun$^{1,2}$, Hongxin Hu$^3$\\\small$^1$Institution for Network Sciences and Cyberspace, Tsinghua University\\$^2$Tsinghua National Laboratory of Information Science and Technology (TNList)\\$^3$Clemson University\\\small mengzl15@mails.tsinghua.edu.cn, junbi@tsinghua.edu.cn, hpwang@seu.edu.cn, c-sun14@mails.tsinghua.edu.cn, hongxih@clemson.edu}
\title{\textsf{CoCo}: Compact and Optimized Consolidation of Modularized Service Function Chains in NFV}
\maketitle

\begin{abstract}
The modularization of Service Function Chains (SFCs) in Network Function Virtualization (NFV) could introduce significant performance overhead and resource efficiency degradation due to introducing frequent packet transfer and consuming much more hardware resources. In response, we exploit the \textit{lightweight} and \textit{individually scalable} features of \textit{elements} in Modularized SFCs (MSFCs) and propose \textsf{CoCo}, a \textit{compact} and \textit{optimized} consolidation framework for MSFC in NFV. \textsf{CoCo} addresses the above problems in two ways. First, \textsf{CoCo} Optimized Placer pays attention to the problem of \textit{which elements to consolidate} and provides a performance-aware placement algorithm to place MSFCs compactly and optimize the global packet transfer cost. Second, \textsf{CoCo} Individual Scaler innovatively introduces a \textit{push-aside scaling up} strategy to avoid degrading performance and taking up new CPU cores. To support MSFC consolidation, \textsf{CoCo} also provides an automatic runtime scheduler to ensure fairness when elements are consolidated on CPU core. Our evaluation results show that \textsf{CoCo} achieves significant performance improvement and efficient resource utilization. 
\end{abstract}

\section{Introduction}
\label{sec:intro}

Network Function Virtualization (NFV)~\cite{11guerzoni2012network} was recently introduced by replacing traditional hardware-based dedicated middleboxes with virtualized Network Functions (vNFs). Compared to the legacy network, NFV brings benefits of easy development, high elasticity, and dynamic management. Meanwhile, network operators often require traffic to pass through multiple vNFs in a particular sequence (e.g. Firewall$\Rightarrow$NAT$\Rightarrow$Load Balancer), which is commonly referred to as a Service Function Chain (SFC)~\cite{82kumar2015service}. To fasten the development of vNFs, many recent research efforts~\cite{ancs2012xomb, sosr2015slick, 92bremler2016openbox, nsdi2012comb} proposed to break traditionally monolithic Network Functions (NFs) into processing \textit{elements}, which could form a Modularized Service Function Chain (MSFC). For example, Intrusion Detection System (IDS) can be broken into a Packet Parser element and a Signature Detector element. In this way, new vNFs could be built based on a library of elements, which could significantly reduce human development hours. 

However, introducing modularization into NFV brings two major drawbacks. First, in NFV networks, each vNF is usually deployed in the form of Virtual Machine (VM) with separated CPU cores and isolated memory resource~\cite{conext2016flurries}. When traversing a SFC, a packet has to be queued and transferred between VMs, which could introduce communication latency~\cite{72hwang2015netvm}. An MSFC requires more times of packet transmission between elements than its corresponding SFC, which may degrade the chain performance. Second, due to modularization, to deploy an MSFC, we need to consume much more (possible 2$\times$ or more) hardware resources to accommodate all processing elements compared with a SFC with monolithic vNFs, which compromises resource efficiency. 

Some research efforts have been devoted to addressing the problems above. OpenBox~\cite{92bremler2016openbox} addressed the performance problem by merging the common elements used by different vNFs in an MSFC to decrease the latency. However, OpenBox is constrained for limited cases where an MSFC comprises repeated elements whose internal rules belonging to different vNFs do not conflict with each other. NFVnice~\cite{sigcomm2017nfvnice} addressed the resource efficiency problem by consolidating several NFs onto a CPU core with containers. However, it was designed at NF-level and ignorant of the new problems of modularization such as frequent inter-VM packet transfer. Also, it did not consider the placement problem of \textit{which elements to consolidate}, which is also significant to improve performance and resource efficiency. Inappropriate consolidation may worsen performance by transferring packets repeatedly. 

At the same time, a closer look into the modularization technique reveals some features of modularization that could benefit both the performance and resource efficiency of MSFCs. Modularization introduces processing elements that are \textit{lightweight} and \textit{individually scalable}~\cite{cloudcom2016microservice,Newman2015microservices}. Therefore, we could consolidate several lightweight elements on the same VM, i.e. CPU core, to reduce hardware resource consumption and improve resource efficiency. Also, by considering which elements to consolidate, performance can be improved by reducing inter-VM packet transfer inside MSFC. Furthermore, in the situation where an element is overloaded, we can only scale out the overloaded lightweight element itself individually instead of its corresponding monolithic NF, which could significantly reduce the scaling cost~\cite{37gember2014opennf}. The scaled out replica can also be consolidated onto an already working VM without consuming an extra CPU core to further save resource.

Therefore, based on the above observations, we propose \textsf{CoCo}, a compact and optimized element consolidation framework to improve the performance and resource utilization efficiency for MSFC in NFV. To the best of our knowledge, \textsf{CoCo} is the first framework addressing the optimal consolidation for MSFC in NFV.
%on \textit{which elements to consolidate}. 
The key idea of \textsf{CoCo} is to reduce inter-VM packet transfer and fully utilize the processing power of CPU by consolidating appropriate elements on the same VM. %As shown in Fig.~\ref{fig:arch}, \textsf{CoCo} framework consists of \textsf{CoCo} controller for performance-aware consolidation placement (Optimized Placer) and elasticity control (Individual Scaler), and \textsf{CoCo} Infrastructure enabling element consolidation. 
However, we encounter three main challenges in our design: 

\begin{itemize}
\item
For SFC placement, to optimize the performance of MSFCs, we are challenged to carefully analyze the cost of inter-VM packet transfer via a virtual switch (vSwitch)~\cite{15openvswitch}. Moreover, we are challenged to design a performance-aware placement algorithm to consolidate appropriate elements together.
%when designing placement algorithm, we need to carefully analyze the inter-CPU packet transfer cost and model the performance-aware placement correctly. 
\item
For SFC elasticity control, careless placing the scaled out replica may introduce additional packet transfer between VMs and %scaling of elements requires state migration and possibly 
frequent state synchronization among different replicas of the element, which may degrade the performance significantly (possibly up to tens of \textit{ms}~\cite{37gember2014opennf}). We are challenged to avoid performance degradation. %To address this challenge, %\textsf{CoCo} exploits the lightweight and individually scalable nature of elements and 
%\textsf{CoCo} proposes an innovative \textit{push-aside element scaling up} mechanism with individual scalability. The overloaded element can \textit{push} its adjacent element \textit{aside} to other VMs and \textit{scales} itself \textit{up} to alleviate overload. 
%Second, for individually scaling, there will be frequent internal states sharing and synchronizing among replicas, which may cause significant performance degradation \cite{37gember2014opennf}. How to avoid potential performance overhead caused by states synchronization and increased packet transferring remains challenging. 
\item
For runtime SFC management, when consolidating multiple elements on the same CPU core, we need to ensure fairness when scheduling CPU resources among elements. However, traditional approach~\cite{sigcomm2017nfvnice} requires \textit{manual} configuration of element priorities, which is time-consuming and lacks scalability. We are challenged to design an automatic scheduler. %In response, \textsf{CoCo} designs an automatic processing speed-aware scheduler. % to address this challenge.
\end{itemize}
%Finally, when consolidating several elements together, traditional load-aware scheduling algorithms are no longer suitable to ensure fairness because different elements have different processing speed. A naive solution is to specify priorities for elements. However, dynamically configuring priorities concerning both varying traffic load and Quality of Service (QoS) is man-in-the-loop for network administrators.

\begin{figure}
\centering
\includegraphics[scale=0.5]{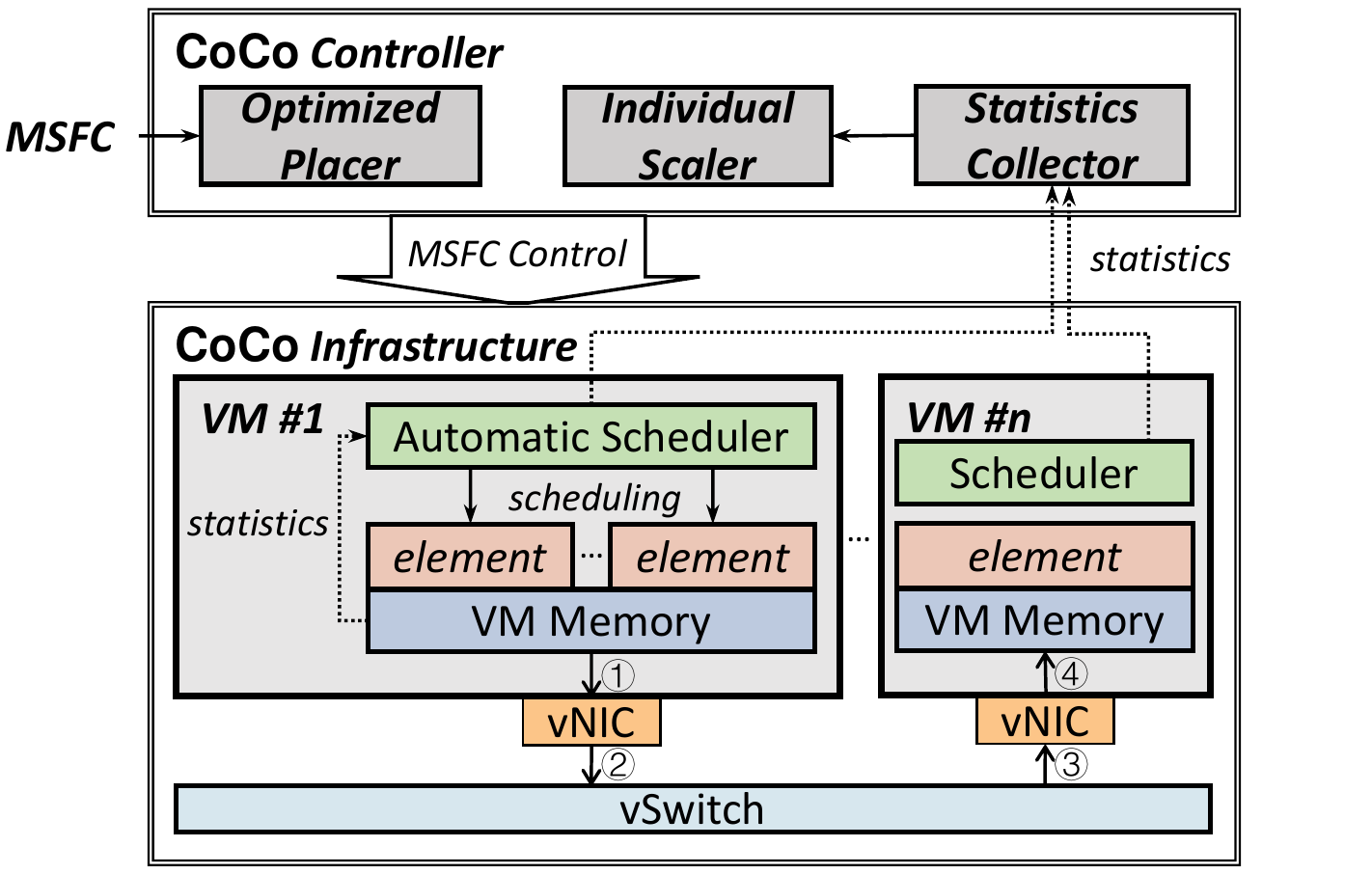}
\caption{\textsf{CoCo} Framework Overview}
\label{fig:arch}
\end{figure}

To address the above challenges, as shown in Fig.~\ref{fig:arch}, we design the \textsf{CoCo} controller for performance-aware consolidation placement (Optimized Placer) and elasticity control (Individual Scaler), and the \textsf{CoCo} Infrastructure that supports automatic resource scheduling. In \textit{Optimized Placer}, \textsf{CoCo} models the performance cost of inter-VM packet transfer and proposes a 0-1 Quadratic Programming-based placement algorithm. In \textit{Individual Scaler}, \textsf{CoCo} proposes an innovative \textit{push-aside element scaling up} strategy as well as a greedy scaling out method for efficient element scaling. %, in which the overloaded element can \textit{push} its adjacent element \textit{aside} to other VMs and \textit{scales} itself \textit{up} to alleviate overload. 
Finally, we design an \textit{Automatic Scheduler} in the \textsf{CoCo} infrastructure that schedules CPU resources based on the processing speed of each element.

In this paper, we make the following contributions:

\begin{itemize}

\item We introduce the problem of which elements to consolidate and model the performance cost of inter-VM packet transfer. We then design an Optimized Placer in \textsf{CoCo} controller and propose a performance-aware placement algorithm to achieve optimal performance of MSFCs. (Section~\ref{sec:place}) 

\item We design an Individual Scaler in \textsf{CoCo} controller for individual scaling of elements. We propose an innovative push-aside scaling up strategy as well as a greedy scaling out method to alleviate the hot spot with little performance and resource overhead. (Section~\ref{sec:scale}) %to avoid performance degradation and a greedy scaling out method to achieve resource efficiency when alleviating the hot spot. 

\item We design a runtime CPU Automatic scheduler in \textsf{CoCo} infrastructure to automatically ensure fairness between multiple elements on the same CPU core with respect to their different processing speed. (Section~\ref{sec:schedule})

\item We evaluate the effectiveness of \textsf{CoCo} framework. Evaluation results show that \textsf{CoCo} could improve both performance and resource efficiency. (Section~\ref{sec:eva})

\end{itemize} 

\section{Performance-aware MSFC Placement}
\label{sec:place}

In this section, we present the \textsf{CoCo} elements placement algorithm inside Optimized Scaler of \textsf{CoCo} Controller for the initial deployment of an MSFC. We have the following goal in mind in our design: \textit{We prefer to consolidate adjacent elements in an MSFC on the same VM and place the MSFC compactly to reduce inter-VM packet transfer cost.} 

In the following, we first analyze the one-hop inter-VM packet transfer cost due to vSwitch-based forwarding. We then find the relationship between CPU utilization and processing speed for an element. These two analyses serve as the fundamentals of placement algorithm of MSFC, which usually contains multiple hops and multiple elements. 

\subsection{Packet Transfer Cost Analysis}
\label{sec:cost}

In NFV implementation, usually elements are implemented as VMs separately with dedicated CPU cores~\cite{conext2016flurries}. To simplify the resource constraint analysis, we assume that each VM is implemented on one CPU core, which could easily be extended to situations where a VM is allocated with multiple CPU cores (Section~\ref{sec:discuss}). When packets are consolidated on the same VM with Docker Container~\cite{96merkel2014docker}, intra-VM packet transferring is simple. With shared memory technique provided by Docker, we can directly deliver the pointer on memory of packet from one element to another with negligible latency (about $3~\mu s$ under our implementation). However, when packets are transferred between VMs, they must go through four steps, as shown in Fig.~\ref{fig:arch}. First, packets are copied from memory to the virtual NIC (vNIC) on the source VM (Step~\circled{1}). Next vNIC transfers packets to vSwitch (Step~\circled{2}). And then packets are delivered reversely from vSwitch to the vNIC of destination VM (Step~\circled{3}) and finally from vNIC to memory (Step~\circled{4}). The total transfer delay (about 1~\textit{ms} in our evaluation) degrades the performance of MSFC significantly.

We use \textit{Delayed Bytes (DB)} to represent the packet transfer cost. Theoretically, $DB$ is constrained by the minimum of element throughput, memory copy rate (Step \circled{1} and \circled{4}), and packet transmission rate (Step \circled{2} and \circled{3}). However, memory copy rate and packet transmission rate ($\sim$10~\textit{GB/s} according to~\cite{tanenbaum2009modern}) are much greater than the SFC throughput (99\% are $<$1~\textit{GB/s} in datacenters~\cite{129roy2015inside}). Thus $DB$ is directly constrained by the throughput between elements. We denote the throughput as $\Theta$ and the additional four-step transfer delay as $t_d$. Thus

\begin{equation}
\label{eq:delay}
DB=\Theta\cdot t_d
\end{equation}

In the placement analysis, we will use the total sum of $DB$ as our optimization target of performance overhead.

\subsection{Resource Analysis}

Before globally optimizing the total sum of $DB$, we need to analyze and find the constraints on CPU resource utilization. For a certain element, as the throughput increases, it will consume more CPU resources to process. Thus for each type of element $i$, we can measure a respective one-to-one mapping function between CPU utilization $r$ and processing speed $v$: 

\begin{equation}
\label{eq:cpu}
r_i=\phi_i(v_i)~\mathrm{and}~v_i=\phi_i^{-1}(r_i)
\end{equation}

%The mapping will be the same for the same type of elements due to the similarity of processing logic. 
In implementation, network administrators can measure the mapping function for each type of element in advance, which will be introduced in Section~\ref{sec:meas}. %, \textsf{CoCo} provides a measurement method by constraining CPU resource and testing the maximum throughput of element. 

Docker-based consolidation technique is lightweight and takes few resources~\cite{conext2016flurries}. Thus, we can directly add up respective CPU utilizations of elements to estimate the total CPU utilization. Also note that Eq.~\ref{eq:cpu} is an upper bound estimation for CPU utilization given throughput. When several elements are consolidated together, the alternate scheduling mechanism by reusing the idle time caused by interrupts~\cite{tanenbaum2009modern} can enable a higher total throughput. 

\subsection{MSFC Placement Algorithm}

\begin{figure}
\centering
\subfigure[Topology 1]{
\includegraphics[scale=0.6]{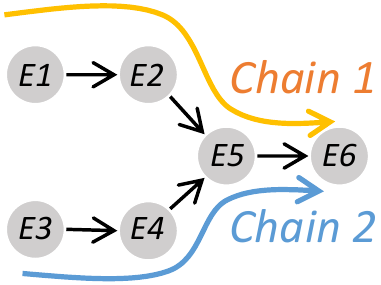}
\label{fig:chain1}
}
\subfigure[Topology 2]{
\includegraphics[scale=0.6]{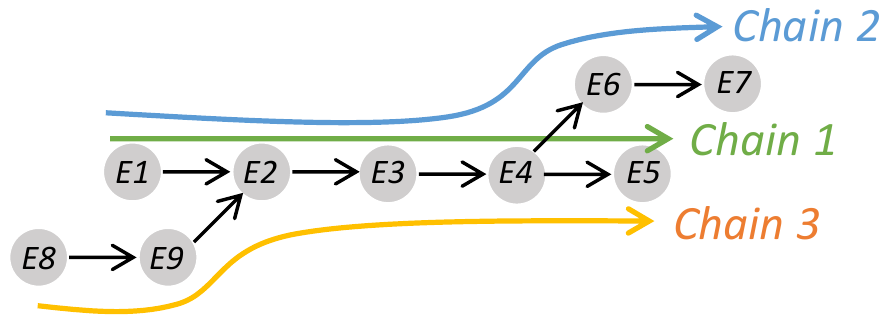}
\label{fig:chain2}
}
\caption{Two Examples of Processing Graph $\mathcal{G}=(\mathcal{V},\mathcal{E})$}
\label{fig:graph}
\end{figure}

We first abstract the packet processing in MSFCs as a directed acyclic processing graph, denoted as $\mathcal{G}=(\mathcal{V},\mathcal{E})$. Each node $k\in\mathcal{V}$ represents an element and each edge in $\mathcal{E}$ represents a hop between elements in an MSFC. $k\in\mathcal{I}$ represents the VMs, i.e. CPU cores. Service chains, denoted as $\mathcal{C}$, are defined by tenants. Fig.~\ref{fig:graph} shows two examples of processing graph. There are two chains and six elements in Fig.~\ref{fig:chain1}. Chain~1 is E1$\Rightarrow$E2$\Rightarrow$E5$\Rightarrow$E6 and Chain~2 is E3$\Rightarrow$E4$\Rightarrow$E5$\Rightarrow$E6.
To consolidate compactly, we assume that the processing speed of each elements on the same chain matches the throughput of the entire chain at initial placement. We denote the throughput of chain $j$ as $\Theta_j$. Conservatively, network administrators can estimate $\Theta_j$ with its required bandwidth according to Service Level Agreement~\cite{luizelli2015piecing}. 

$\alpha_i^j\in\{0,1\}$ indicates whether element $i$ is on chain $j$. $\pi_i^j$ represents the upstream element of element $i$ on chain $j$, which can be realized with a doubly linked list. To ensure robustness, when $i$ is the first element on chain $j$, we set $\pi_i^j=i$. When $\alpha_i^j=0$, we set $\pi_i^j=0$. %For example, in Fig.~\ref{fig:chain1}, we have $\pi_5^1=2$, $\pi_6^1=\pi_6^2=5$ and $\alpha_4^1=0$. 

\textsf{CoCo} applies a 0-1 Integer Programming (0-1 IP) algorithm to minimize the inter-VM overhead. $x_{i,k}$ is a binary indicator of whether placing element $i$ onto VM $k$. For chain $j$, we analyze the performance overhead between each element $i$ and its upstream element $\pi_i^j$ on it. From Eq.~\ref{eq:delay}, %the $DB$ of one hop on chain $j$, denoted as $DB_j$, keeps unchanged anywhere on the chain since the estimated throughput of chain $j$ is invariant. T
the hop from $\pi_i^j$ to $i$ will incur an inter-VM cost of $DB_j$ if and only if they are not placed on the same VM, i.e.
\[
x_{i,k} x_{\pi _i^j,k}=0,~\forall k\in \mathcal{I}
\]
Thus we can use $\left(1-\sum_{k\in\mathcal{I}} x_{i,k} x_{\pi_i^j,k}\right)\in\{0,1\}$ to indicate whether element $\pi_i^j$ and $i$ are consolidated together. Then we add up $DB$ of all inter-VM hops in chain $j$ and further add up $DB$ of different chains as our objective function. \textsf{CoCo} aims at minimizing the total inter-VM cost to improve performance: 

\begin{equation}
\label{eq:total}
\min~\sum_{j\in\mathcal{C}}\sum_{i\in \mathcal{V}}\alpha_i^j DB_j
\left(1-\sum_{k\in\mathcal{I}} x_{i,k} x_{\pi_i^j,k}\right)
\end{equation}

Meanwhile, the following constraints should be satisfied:

(1) $x_{i,k}\in\{0,1\},~\forall i\in \mathcal{V},~k \in \mathcal{I}$

//An element is either consolidated on VM $k$ or not.

(2) $\sum_{k\in \mathcal{I}} x_{i,k} =1, ~\forall i\in \mathcal{V}$

//An element can only be placed onto one VM.

(3) $\sum_{i\in\mathcal{V}}\left[x_{i,k}\cdot \phi_i\left(\sum_{j\in \mathcal{C}}\alpha_i^j\Theta_j\right)\right]\leqslant 1,~\forall k\in \mathcal{I}$

//Each CPU core cannot be overloaded at initial placement.

Note that if the estimated through of element $i_0$ is so high that it cannot be placed on one CPU core, i.e. 

\begin{equation}
\label{eq:overload}
\exists i_0\in \mathcal{V},~s.t.~
\sum_{j\in \mathcal{C}}\alpha_{i_0}^j\Theta_j>\phi_{i_0}^{-1}(1)
\end{equation}
constraint (2) and (3) may conflict. This is due to the incorrect orchestration between flows and elements. Actually it rarely happens in the real world and never happens in our evaluation. In this situation, scaling out is needed. For overloaded element $i_0$, if there is only one chain $j_0$ containing it, \textsf{CoCo} scales out $\lceil\frac{\Theta_{j_0}}{\phi_{i_0}^{-1}(1)}\rceil$ replicas and performs load-balancing among them. If there are several chains containing $i_0$ (such as E5 in Fig.~\ref{fig:chain1}), \textsf{CoCo} scales it out to multiple replicas based on a knapsack algorithm on chain. Then \textsf{CoCo} splits the overlapped chains to different replicas and reconstructs processing graph $\mathcal{G}'$ to ensure that Eq.~\ref{eq:overload} does not hold for $\forall i\in \mathcal{G}'$. 

From the analysis above, we find that Eq.~\ref{eq:total} is a 0-1 Quadratic Programming problem and can be solved within limited time and space~\cite{quadratic-programming}. By solving the above formulations, we can get the performance-aware optimized placement solution. We evaluate this algorithm in Section~\ref{sec:eva-place}.

\section{Optimized Individual Scaling}
\label{sec:scale}

\begin{figure}
\centering
\subfigure[Monolithic scaling of IDS]{
\label{fig:monolithic}
\includegraphics[scale=0.55]{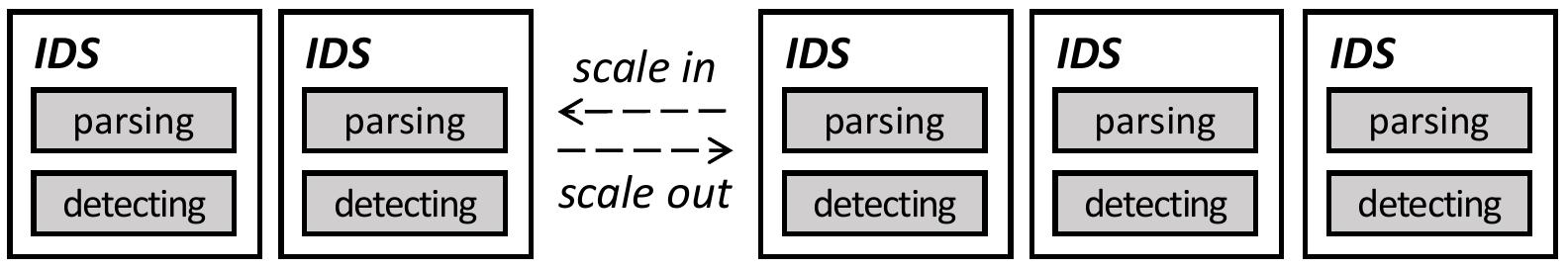}
}
\subfigure[Individual scaling of modularized IDS]{
\label{fig:individual}
\includegraphics[scale=0.55]{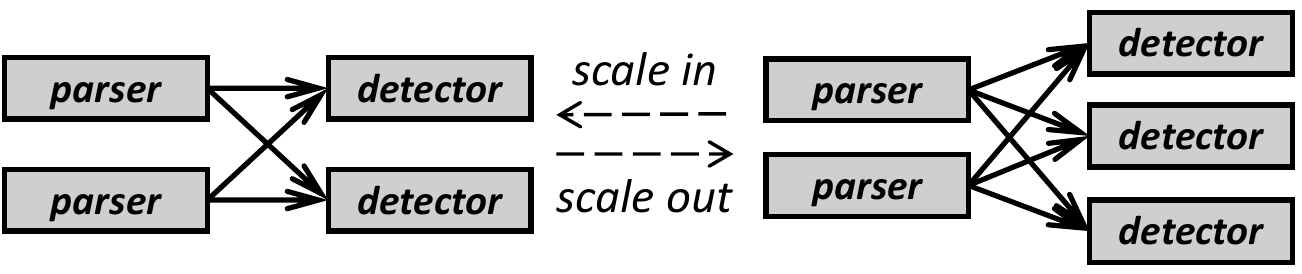}
}
\caption{Individually scalability}
\label{fig:scale}
\end{figure}

\begin{figure*}
\centering
\subfigure[MSFC before scaling]{
\label{fig:before-scale}
\includegraphics[scale=0.47]{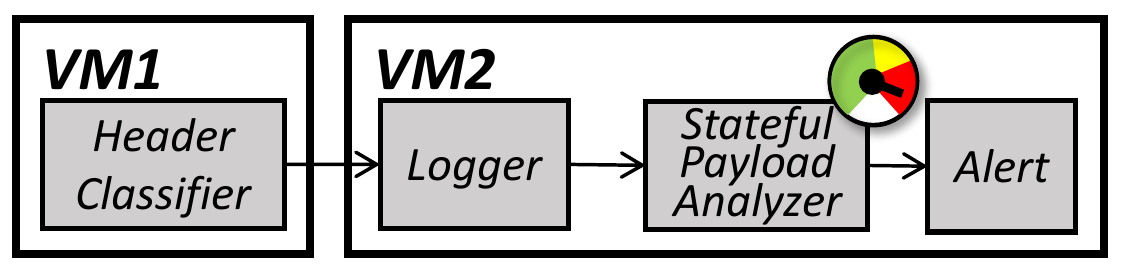}
}
\subfigure[Scaling out with traditional method]{
\label{fig:scale-out}
\includegraphics[scale=0.47]{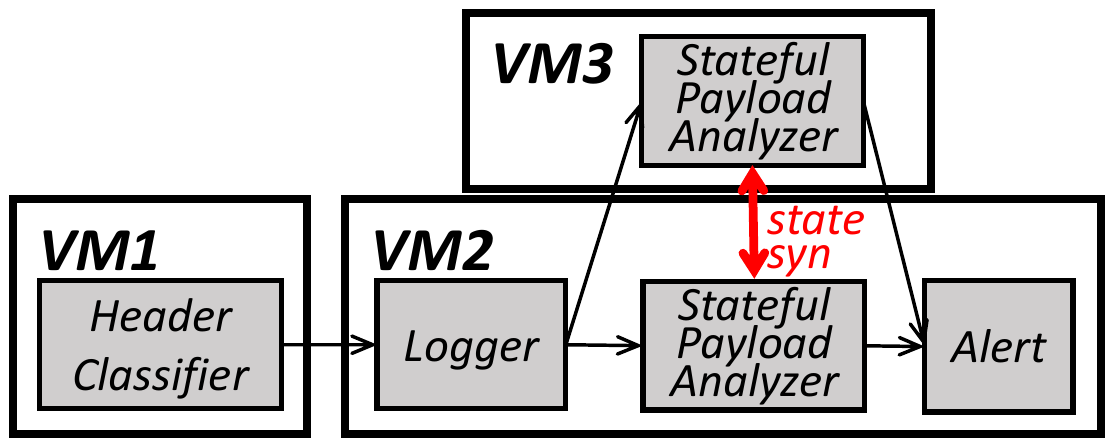}
}
\subfigure[Push-aside scaling up]{
\label{fig:scale-up}
\includegraphics[scale=0.47]{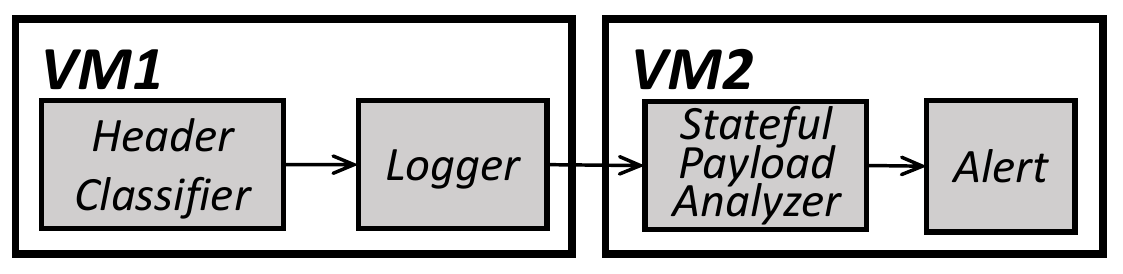}
}
\caption{Comparison of push-aside scaling up and traditional scaling out}
\label{fig:aside}
\end{figure*}

For monolithic NFs, all components must be scaled out at the same time when overloaded, which will take up more resources than needed. Also, continuously synchronizing numerous internal states will introduce significant overhead~\cite{37gember2014opennf}. After modularization, when traffic increases, only the overloaded elements need to scale out. For example, when the \textit{detector} element of IDS is overloaded, instead of scaling out the whole IDS (Fig.~\ref{fig:monolithic}), we can only scale out the detector element itself (Fig.~\ref{fig:individual}). %In this way, resource-efficiency and flexibility can be achieved. %Also, as the implementation of element are more lightweight than monolithic NFs, the scaled out replica can be consolidated on other already existed VMs. 

However, deciding where to place the scaled out replica is also important. Careless placement will degrade performance and resource efficiency. In this section, \textsf{CoCo} Individual Scaler provides two innovative scaling strategy including performance-aware push-aside scaling up, and resource-aware greedy scaling out, to efficiently alleviate the overload situation. Push-aside scaling up can avoid the performance degradation caused by additional inter-VM packet transfer. Greedy scaling out can achieve resource efficiency by placing replicas on existing VMs. 

\subsection{Push-aside Scaling Up}

When scaling out, the traditional NF-level scaling method taken by~\cite{37gember2014opennf, 131wang2016transparent} simply starts a new VM with taking a new CPU core and scales out the overloaded element to the new VM. %Thus when an element is overloaded, similar to traditional methods at NF-level, a naive method is to scale it out to a newly started VM. 
For example, when the Stateful Payload Analyzer in VM2 is overloaded (Fig.~\ref{fig:before-scale}), traditional method starts VM3 and copies the element to it (Fig.~\ref{fig:scale-out}). However, this will introduce additional latency overhead due to inter-VM packet transfer. For example, in Fig.~\ref{fig:scale-out}, a part of packets will suffer 3 inter-VM hops to go through the total MSFC (VM1$\Rightarrow$VM2$\Rightarrow$VM3$\Rightarrow$VM2). Also, frequent state synchronization between replicas will also degrade the performance.%In this case, frequent state synchronization and increased inter-VM hops still degrade the performance of MSFC. 

However, when an element in an MSFC is overloaded, the VMs that its upstream or downstream elements placed on may be underloaded. Enabled by the lightweight feature of element, we can re-balance the placement of elements on the two VMs. Thus, the key idea of push-aside scaling up is that the overloaded element can \textit{push} its downstream/upstream element \textit{aside} to its downstream/upstream VM and \textit{scale} itself \textit{up} to alleviate the overload. As for Fig.~\ref{fig:before-scale}, we can \textit{migrate} Logger to VM1 and release its CPU resource (Fig.~\ref{fig:scale-up}). By allocating the newly released resource to Stateful Payload Analyzer and scaling it up, the overload can be alleviated.

Push-aside scaling up has two advantages. First, compared to the traditional method, it does not create new inter-VM hops, thus there is no additional packet transfer cost. Second, it does not create a new replica but allocates more resources to overloaded element. Thus push-aside scaling up does not suffer the state share and synchronization problems~\cite{37gember2014opennf}. 

The algorithm includes the following four steps. 

\noindent\textbf{Step 1: Check the practicability.} If the estimated throughput $\Theta_{i_0}^{exp}$ for element $i_0$ is so large that the overload on $i_0$ cannot be alleviated with one CPU core, i.e. $\Theta_{i_0}^{exp}>\phi_{i_0}^{-1}(1)$, push-aside scaling up will not work. This happens only when an extremely large burst comes. In this situation, algorithm terminates and \textsf{CoCo} goes to greedy scaling out. 

\noindent\textbf{Step 2: Find border elements.} For element $i_0$ in chain $j_0$, \textsf{CoCo} first finds out its upstream and downstream border elements $up\_border_{i_0}^{j_0}$ and $down\_border_{i_0}^{j_0}$. Upstream border element refers to the element that $up\_border_{i}^j$ and $i$ are placed in the same VM but $up\_border_{i}^j$ and $\pi_{up\_border_{i}^j}^j$ are placed separately. %In Fig.~\ref{fig:push-before}, for the only chain 1, $up\_border_{4}^1=3$ and $down\_border_{1}^1=2$. 
With doubly linked list, \textsf{CoCo} goes through elements hop by hop to find out border elements and composes them into a set $\mathcal{B}_{i_0}$. If $i_0$ is contained by several chains, \textsf{CoCo} checks each chain and composes the results into $\mathcal{B}_{i_0}$. 

\noindent\textbf{Step 3: Check whether it can be migrated.} After finding out the border elements %in $\mathcal{B}_{i_0}$
, \textsf{CoCo} checks whether they can be migrated to the adjacent VM. Suppose both $b_0\in\mathcal{B}_{i_0}$ and $i_0$ are on chain $j_0$. We denote the adjacent VM of $b_0$ as $k^{adj}_{b_0}$. If
\begin{equation}
\phi_{b_0}(\Theta_{j_0})+
\sum_{i\in\mathcal{V}}x_{i,k^{adj}_{b_0}}\cdot\phi_i
\left(\sum_{j\in\mathcal{C}}\alpha_i^j\Theta_j\right)<1
\end{equation}
which means there is available resource for $b_0$ on $k^{adj}_{b_0}$, we can migrate $b_0$ to $k^{adj}_{b_0}$ and release its resource. Similarly, we can check all $b\in\mathcal{B}_{i_0}$ and its respective $k^{adj}_b$. If none can be migrated, %push-aside scaling up terminates and 
\textsf{CoCo} goes to the greedy scaling out strategy. This happens only when all of the adjacent VMs of $i_0$ do not have enough resource, which in practice rarely happens. If some of them can be migrated, \textsf{CoCo} composes them into $\mathcal{B}'_{i_0}\subset\mathcal{B}_{i_0}$.

\noindent\textbf{Step 4: Check whether overload can be alleviated.} At last, \textsf{CoCo} checks if migrating all elements in $\mathcal{B}'_{i_0}$ will make enough room for $i_0$ to scale up to alleviate the overload. Otherwise the migration will be useless. \textsf{CoCo} calculates the needed resource $r_{i_0}^*=\phi_{i_0}(\Theta_{i_0}^{exp})-\phi_{i_0}(\Theta_{i_0}^{cur})$, where $\Theta_{i_0}^{cur}$ is the current processing speed of $i_0$. The CPU utilization of element $b'\in\mathcal{B}'_{i_0}$ satisfies
$r_{b'}=\phi_{b'}\left(\sum_{j\in\mathcal{C}}\alpha_{b'}^j\Theta_j\right)$. 
If 
$\sum_{b'\in\mathcal{B}'_{i_0}} r_{b'}<r_{i_0}^*$, 
which means migration cannot release enough resource, the algorithm terminates and \textsf{CoCo} goes to greedy scaling out. Else, push-aside scaling up can be applied. Also, aware that migrating elements from one VM to another has performance cost and controller overhead \cite{37gember2014opennf}, \textsf{CoCo} tries to minimizes the number of elements to migrate. \textsf{CoCo} finds a subset $\mathcal{B}''_{i_0}\subset\mathcal{B}'_{i_0}$ with minimal number of elements that satisfies $\sum_{b''\in\mathcal{B}''_{i_0}}r_{b''}\geqslant r^*$. 

Moreover, to avoid potential frequently scaling up among elements, network administrators can set a timeout between each time of scaling up. In this way, we can alleviate the overload with minimum elements to migrate. 

\subsection{Greedy Scaling Out}

If the overloaded element can push none of border elements aside to other VMs, Individual Scaler have to scale it out to somewhere else. In this situation, performance degradation caused by scaling out is unavoidable. Even so, we can still save resource by placing the new replica to an already working VM. 

\textsf{CoCo} decides the VM to place the replica based on a greedy algorithm. First, it calculates the remained resource of each VM and sorts them in increasing order. Next, \textsf{CoCo} greedily compares it with the needed resource $r_{i_0}^*$. When the remained resource of any VM, i.e. CPU core, is larger than $r_{i_0}^*$, \textsf{CoCo} places the replica there. If none of the VMs have available CPU resource, \textsf{CoCo} will call up new VMs and specify new CPU cores, just as the traditional method does. 

\section{Automatic Consolidation Scheduling}
\label{sec:schedule}

When consolidating several elements onto one CPU core, \textsf{CoCo} uses one Docker \cite{96merkel2014docker} for each element. At this time, we need a scheduling algorithm to enable fair resource allocation. However, as we discussed above, traditional rate-proportional scheduling methods \cite{ton1998scheduling} and priority-aware scheduling methods \cite{sigcomm2017nfvnice} are not scalable due to massive manual configurations. Thus, we design a novel scheduling algorithm to match processing speed of elements with its throughput to automatically achieve both fairness and efficiency. Here, we take CPU resource as the allocation variable since CPU is more likely to become a bottleneck resource than memory~\cite{sigcomm2017nfvnice}, especially when elements are densely consolidated in MSFC. 

\textsf{CoCo} takes an incrementally adaptive adjustment scheduling algorithm. The algorithm tries to match the processing speed of each element with its packet arrival rate. It incrementally adjusts the CPU utilization of the next scheduling period based on the statistics of current scheduling period. The detailed algorithm is introduced below.

In consolidation, CPU resources are scheduled among elements by periodically allocating \textit{time slices} with CGroup~\cite{linux-container}. Suppose scheduling period is $T$. For element $i$ on a VM, we can get its CPU utilization proportion $r_i$ by counting the number of time slices. Note that $r_i$ is a proportion thus $\sum_i r_i =1$. Also, we can get current buffer size $B_i$ and last time buffer size $B_i'$. From Eq.~\ref{eq:cpu}, we can know the processing speed $v_i$ satisfies $v_i=\phi_i^{-1}(r_i)$. We denote $B^*_i, v^*_i$ and $r^*_i$ as the \textit{predicted} buffer size, processing speed and CPU utilization at the next scheduling period. 

Our scheduling algorithm is based on the matching principle: \textit{For all elements, their buffer variations should be proportional to respective processing speeds}, i.e.
\begin{equation}
\label{eq:schedule}
\frac{B^*_i-B_i}{v^*_i T}=C,~\forall i \in \{1,\cdots,n\}
\end{equation}

By modeling in this way, we try to ensure fairness among elements and effectively allocate resources. The key idea is to match the processing speed with its respective packet arrival rates. In this way, the element with a lower processing speed and smaller flows can also attain an appropriate proportion of CPU. However, when several elements are consolidated together, downstream element may directly read the packet that are already loaded into memory by its upstream element. Thus we cannot get the actual packet arrival rate by simply measuring at the last switch. Naively adding statistics measuring module on the top of Docker will introduce unnecessary overhead. Instead, we can infer the arrival rate $v_{ai}$ from the variation of buffer size, i.e.
\begin{equation}
\label{eq:arr_rate}
v_{ai}=\frac{B_i-B_i'}{T}+v_i
\end{equation}

As traffic usually does not vary sharply \cite{145benson2011microte}, we can assume that in a scheduling period (usually at millisecond level), the packet arrival rate keeps invariant, i.e. $v_{ai}=v^*_{ai}$. By substituting $B^*_i$ in Eq.~\ref{eq:schedule} with Eq.~\ref{eq:arr_rate}, we can get the CPU proportion for element $i$ in the next scheduling period:
\begin{equation}
r_i^*=\phi_i(v_i^*)=\phi_i\left(\frac{\frac{B_i-B_i'}{T}+v_i}{C+1}\right)
\end{equation}

The sum of CPU utilization needs to be normalized, thus $C$ subjects to
\begin{equation}
\label{eq:const}
\sum_i \phi_i
\left(\frac{\frac{B_i-B_i'}{T}+v_i}{C+1}\right)=1
\end{equation}

Although we cannot get an explicit expression of $C$, we can first randomly specify an initial value for $C$ and then normalize $r_i^*$. Comparing to the millisecond level scheduling period, the solving time in this way is negligible. 

Another important function of consolidation scheduler is to tell the individual scaler when to scale out. A direct indicator is the buffer size of an element. If buffer is overflowed, packet loss incurs and the element is definitely overloaded. At this time, scheduler can do nothing but execute the scaling up or scaling out methods, as introduced in Section~\ref{sec:scale}.

\section{Preliminary Evaluation}
\label{sec:eva}
In this section, we first introduce our methods on measuring throughput-CPU utilization mapping function. Then we build \textsf{CoCo} with Docker~\cite{96merkel2014docker} to consolidate elements on VMs, and enable inter-VM packet forwarding with Open vSwitch (OVS)~\cite{15openvswitch}. We take the low-level dynamical element migration mechanism from OpenNF%\footnote{We obtain the source code of OpenNF from http://opennf.cs.wisc.edu/code.}
 and evaluate the effectiveness of high-level \textit{push-aside scaling up} strategy compared to OpenNF~\cite{37gember2014opennf}. Finally, we evaluate the \textsf{CoCo} performance-aware MSFC placement algorithm based on our simulations. 

We evaluate \textsf{CoCo} based on a testbed with one server equipped with two Intel\circledR~Xeon\circledR~E5-2690 v2 CPUs (3.00GHz, 8 physical cores), 256G RAM, and two 10G NICs. The server runs Linux kernel 4.4.0-31. %We implement 16 VMs separately on 16 physical cores. We use one VM to run the \textsf{CoCo} Manager, one VM for
%, and other VMs for elements. 

\subsection{Throughput-CPU Utilization Mapping}
\label{sec:meas}

\begin{figure}
\centering
\includegraphics[scale=0.6]{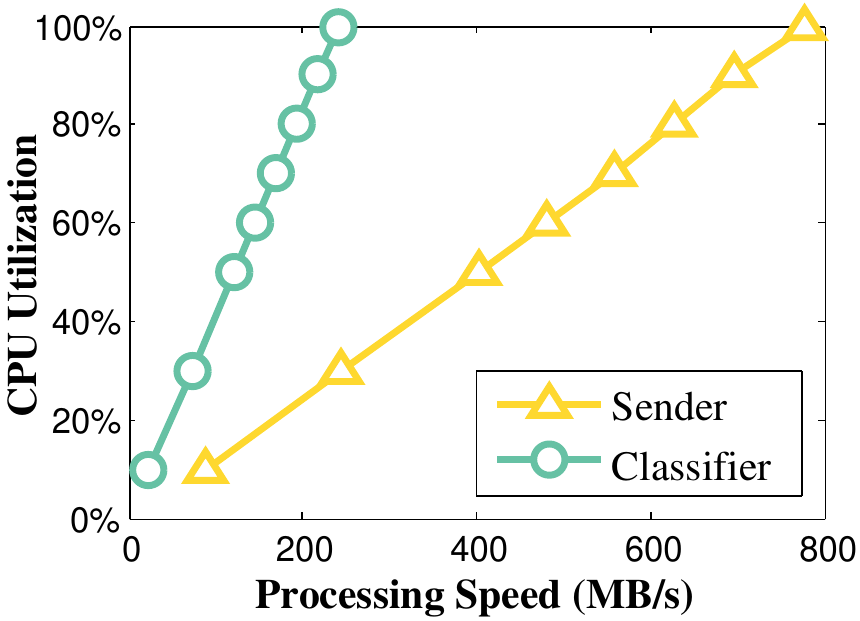}
\caption{Throughput-CPU Utilization Mapping}
\label{fig:cpu}
\end{figure}

To measure the throughput-CPU utilization mapping, we constrain the available CPU utilization for the element by fixing the \texttt{cpu-period} and changing the \texttt{cpu-quota} parameter in Docker. We use a packet sender to test the maximum throughput under the current limited CPU proportion. With this method, administrators can get the mapping function $\phi(v)$.% and continue with placement, scaling and scheduling. 

We measure two types of element with different complexity. Packet Sender represents elements with simple processing logic. IP Address-based Header Classifier contains 100 rules and represents relatively complicated elements. The mapping functions are shown in Fig.~\ref{fig:cpu}. Surprisingly, a strong linearity correlation can be observed. We present the linear regressions of the two mapping functions:
\begin{itemize}
\item \textbf{Sender:} $r=-0.022+0.0013\times v,~R^2=0.9997$
\item \textbf{Classifier:} $r=0.00048+0.0042\times v,~R^2=0.9999997$
\end{itemize}

$r\in[0,1]$ is the CPU utilization and $v$ is the processing speed in \textit{MB/s}. $R^2$ is a measure of goodness of fit with a value of 1 denoting a perfect fit. %Above regression expressions demonstrate a strong linear correlation. 
Thus in practice, we can further simplify the solving procedure by substituting $\phi(v)$ and $\phi^{-1}(r)$ with their linear approximations. 

\subsection{Push-aside Scaling Up}

\begin{figure}
\centering
\includegraphics[scale=0.25]{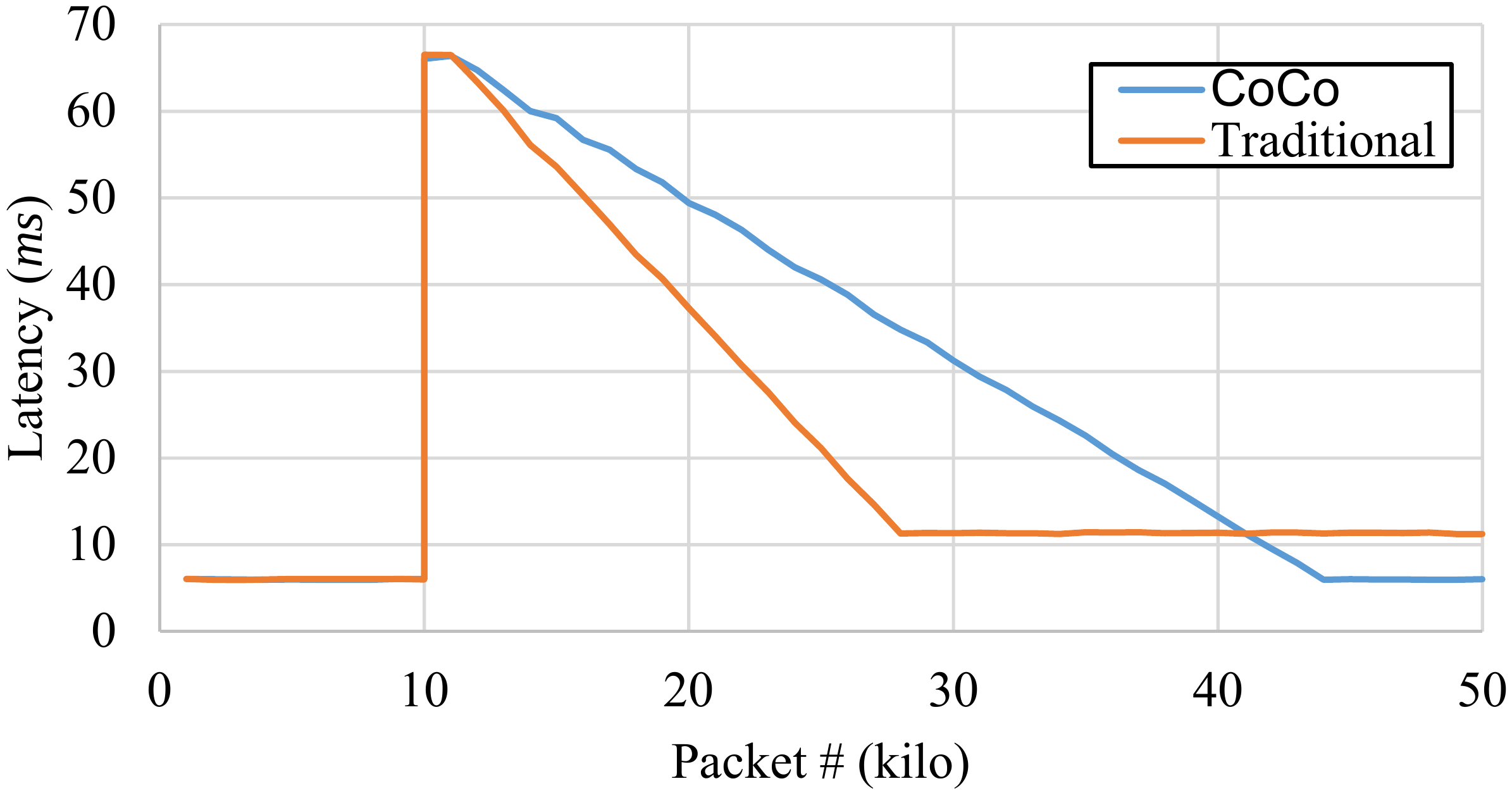}
\caption{Real-time Latency per Packet}
\label{fig:latency}
\end{figure}

To evaluate the performance of push-aside scaling up, we use the MSFC in Fig.~\ref{fig:before-scale}. At first, the throughput of MSFC is 100~kpps, with each packet 512~B. At the 10k-th packet, we increase the traffic to 150~kpps, which causes the Stateful Payload Analyzer overloaded. The traditional method taken by OpenNF~\cite{37gember2014opennf} naively scales out by copying Stateful Payload Analyzer to a newly started VM, as shown in Fig.~\ref{fig:scale-out}. In contrast, \textsf{CoCo} migrates Logger to VM1 and allocate the released resource to Stateful Payload Analyzer. 

For performance, the comparison of two methods on real-time latency of each packet is shown in Fig.~\ref{fig:latency}. The latency at 10k-th packet increases sharply due to the element migration. Traditional method converges a little faster because it consumes more resources and has a higher processing speed. However, \textsf{CoCo} has a lower converged latency of 6~\textit{ms} compared to 11~\textit{ms} of the traditional method. The improvement is achieved by reducing the additional packet transfer and state synchronization latency. % about 13~\textit{ms}, as shown in Fig.~\ref{fig:scale-out}. At the same time, a part of packets are transferred to VM3 and back to VM2, which increases the transfer latency. 
For resource efficiency, with traditional method, the scaled MSFC is allocated with more resources (3 VMs in total). In contrast, 2 VMs are enough with \textsf{CoCo} in this situation (Fig.~\ref{fig:scale-up}) by push-aside scaling up. \textsf{CoCo} achieves a higher resource efficiency by 1.5$\times$. 

\subsection{Performance-aware MSFC Placement}
\label{sec:eva-place}

As for evaluating the performance of placement algorithm, we evaluate the total \textit{DB} in the processing graph. We use an Optimization Toolbox~\cite{matlab-optimization} to solve the 0-1 Quadratic Programming. For large-scale and complicated topologies, quadratic programming can be efficiently solved with some dedicated commercial solvers such as IBM\circledR~CPLEX~\cite{cplex}. We use two topologies shown in Fig.~\ref{fig:graph} and implement all elements as classifiers described in Fig.~\ref{fig:cpu}. We randomly select flows from the LBNL/ICSI enterprise trace~\cite{143enterprise} to different chains and repeat the experiment for 1000 times to eliminate the randomness. We try to place Topology 1 (Fig.~\ref{fig:chain1}) on 2 VMs and Topology 2 (Fig.~\ref{fig:chain2}) on 4 VMs. 

Since that there is no ready-made solution on \textit{which elements to consolidate}, we compare \textsf{CoCo} with two strawman solutions, including a \textit{greedy} mechanism, which greedily places elements onto VMs chain by chain, and a \textit{random} mechanism, which randomly selects available VMs to place elements. 

\begin{figure}
\centering
\includegraphics[scale=0.22]{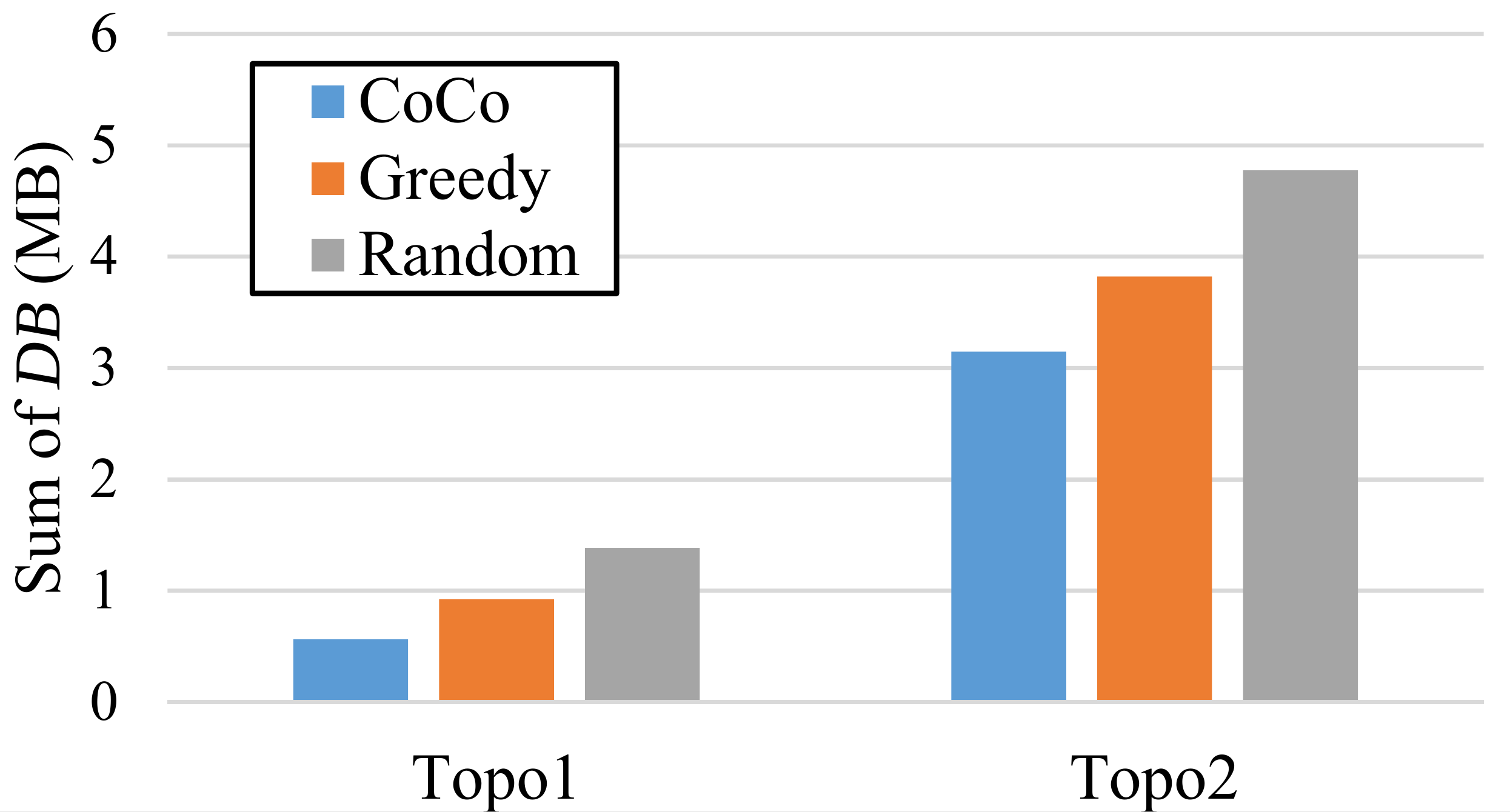}
\caption{Sum of Delayed Bytes}
\label{fig:cost}
\end{figure}

An important feature of \textsf{CoCo} placement is performance-aware, which is evaluated as the sum of \textit{DB}. The total \textit{DB} in processing graph of successful placements with different strategies is shown in Fig.~\ref{fig:cost}. For Topo1, \textsf{CoCo} reduces the total \textit{DB} by 2.46$\times$ compared to random strategy and by 1.64$\times$ compared to greedy strategy. For Topo2, even the lengths of chains increase and more inter-VM packet transfers are unavoidable, \textsf{CoCo} still outperforms the random strategy by 1.52$\times$ and greedy strategy by 1.21$\times$. 

\begin{table}[t]
\centering
\caption{Placement Failure Rate (Of 1000 Tests)}
\label{tabl:success_place}
\begin{tabular}{|c|c|c|c|}
\hline
 & \textsf{CoCo} & Greedy & Random \\
 \hline
Topo1 & 114 (11.4\% fails) & 155 (15.5\% fails) & 214 (21.4\% fails) \\
\hline
Topo2 & 58 (5.8\% fails) & 65 (6.5\% fails) & 78 (7.8\% fails) \\
\hline
\end{tabular}
\end{table}

Another feature of \textsf{CoCo} is resource efficient by compact placement, which can be interpreted as: \textit{Given limited CPU cores, i.e. VMs, for different traffic, \textsf{CoCo} has a higher probability to place all of them on successfully.} As shown in Table~\ref{tabl:success_place}, when placing Topo1, \textsf{CoCo} improves the failure rate by 1.88$\times$ compared to random strategy and 1.36$\times$ compared to greedy strategy of 1000 tests. Note that the placement of Topo1 ($\frac{6~\mathrm{elements}}{2~\mathrm{VMs}}=3$) is tighter than Topo2 ($\frac{9~\mathrm{elements}}{4~\mathrm{VMs}}=2.25$), thus placing Topo1 has a higher failure rate than Topo2. Even so, \textsf{CoCo} improves the failure rate from 7.8\% (random) and 6.5\% (greedy) to 5.8\% for Topo2. 

\section{Related Work}
\label{sec:work}

In this section, we summarize some related work and compare them with \textsf{CoCo}. 

\noindent\textbf{Modularization.} Click \cite{sosp1999click} proposed the idea of modularization and applies it to routers. Recently, Slick~\cite{sosr2015slick} and OpenBox~\cite{92bremler2016openbox} were proposed to detailedly discuss modularized NFs and decouple control plane and data plane of modularized NFs for easy management. Besides, OpenBox focused on merging elements to shorten the processing path length. %We take the abstraction of packet processing from OpenBox. 
However, above works mainly focus on orchestration-level module management and are orthogonal to our optimizations on performance-aware placement and dynamically scaling. 

\noindent\textbf{Consolidation.} CoMb \cite{nsdi2012comb} designed a detailed mechanism to consolidate middleboxes together to reduce provisioning cost. Furthermore, Flurries \cite{conext2016flurries} and NFVnice \cite{sigcomm2017nfvnice} were proposed to share CPU cores among different NFs with the technique of Docker Container \cite{96merkel2014docker}. By modifying Linux scheduling methods, they achieved almost no loss in NF sharing. However, they operated on monolithic NF level and did not consider the problem of which elements (NFs) to consolidate. However, their development details and infrastructure designs could complement our work as the low-level implementation. 

\noindent\textbf{Placement.} There are a lot of research on NF placement in NFV, such as \cite{luizelli2015piecing, mehraghdam2014specifying, savi2015impact}, which focus on the trade-off between traffic load, QoS, forwarding latencies, and link capacities. However, they focused on the placement at NF-level instead of element-level, and thus did not benefit from the lightweight feature of NF modules. Slick~\cite{sosr2015slick} considered placement at element-level. However, all of the work above addressed how to place \textit{middleboxes (vNFs)} onto different \textit{servers} considering complicated and limited physical links. In contrast, \textsf{CoCo} pays attention to the placement problem of \textit{how to consolidate elements} onto different \textit{VMs}, i.e. \textit{cores}. 

\section{Discussions}
\label{sec:discuss}

In this section, we discuss how to extend \textsf{CoCo} and highlight several open issues as future directions.

\noindent\textbf{Multi-core Placement Analysis.} For simplicity, \textsf{CoCo} assumes that each VM is allocated with one CPU core when optimizing placement to satisfy the general applications. In some cases, when tenants allocate multiple CPU cores to a VM, \textsf{CoCo} can be easily extended by considering the resource constraint of multiple CPU cores instead of a single core.%a single-core resource constraints to resource pool constraints.% and considering the hardware resource globally. 

\noindent\textbf{Intra-core Analysis.} \textsf{CoCo} analyzes the inter-core cost caused by vSwitch-based packet transfer. As our future work, by designing cache replacement policies, we may reduce the miss rate of Layer~1 and 2 Cache and further reduce repeatedly packet loading from memory to cache. Moreover, more designs are needed to ensure isolation between consolidated elements. However, those analyses are infrastructure-dependent and differs on various types of CPU, which is beyond our scope. \textsf{CoCo} can be easily extended to analyze intra-core situations on a certain type of CPU. 

\section{Conclusion}
\label{sec:concl}

This paper presents \textsf{CoCo}, a high performance and efficient resource management framework, for providing compact and optimized element consolidation in MSFC. \textsf{CoCo} addresses the problem of which elements to consolidate in the first place and provides a performance-aware placement algorithm based on 0-1 Quadratic Programming. \textsf{CoCo} also innovatively proposes a push-aside scaling up strategy to avoid performance degradation in scaling. \textsf{CoCo} further designs an automatic CPU scheduler aware of the difference of processing speed between elements. Our preliminary evaluation results show that \textsf{CoCo} could reduce packet transfer cost by up to 2.46$\times$ and improve performance at scaling by 45.5\% with more efficient resource utilization. 

\section{Acknowledgement}

We thank anonymous ICC reviewers for their valuable comments. We also thank Zhilong Zheng and Heng Yu from Tsinghua University for providing suggestions on system design and implementation. This work is supported by National Key Research and Development Plan of China (2017YFB0801701), the National Natural Science Foundation of China (No. 61472213) and National Research Foundation of Korea (NRF-2014K1A1A2064649). Jun Bi is the corresponding author.

%\vspace{-0.5\baselineskip}
\bibliography{bibfile}
\bibliographystyle{plain}
\end{document}